\documentstyle[aps,prl,psfig]{revtex}
\begin{document}
\draft
\twocolumn[\hsize\textwidth\columnwidth\hsize\csname
@twocolumnfalse\endcsname
\title{Vortex avalanches and magnetic flux fragmentation in superconductors}
\author{Igor Aranson$^{(1)}$, Alex Gurevich$^{(2)}$, and Valerii Vinokur$^{(1)}$} 
\address{$^{(1)}$
Materials Science Division, Argonne National Laboratory, Argonne, Illinois 60439\\
$^{(2)}$ Applied Superconductivity Center, University of Wisconsin, Madison,
Wisconsin 53706} 
\date{\today}
\maketitle
\begin{abstract}
We report results of numerical 
simulations of  non isothermal dendritic flux penetration  
in type-II superconductors. We propose a generic mechanism of dynamic
branching of a propagating hotspot of a flux flow/normal 
state triggered by a local heat pulse. The branching  occurs when
the flux hotspot reflects from inhomogeneities or the boundary on which magnetization currents 
either vanish, or change direction. Then the hotspot undergoes a cascade of 
successive splittings, giving rise to a 
dissipative dendritic-type flux structure. 
This dynamic state eventually cools down,
turning into a frozen multi-filamentary 
pattern of magnetization currents. 
\end{abstract}
]
\narrowtext

The formation of a macroscopic current-carrying 
critical state in type II  superconductors occurs via penetration of the magnetic flux 
front of pinned vortices from the surface of the sample.  
This process is controlled by a highly nonlinear electric field-current density $(E-J)$
characteristics $E(J,T,B)$, which together with the Maxwell equations 
determine macroscopic electrodynamics of superconductors. \cite{blw,blat,ehb} 
Propagating magnetic flux causes Joule heating, giving rise to global flux jumps and 
thermal quench instabilities which 
are crucial for stable operation of  current-carrying superconductors \cite{fj,gm}. 
Recent advances in the magneto-optical imaging have revealed 
a new class of instabilities of the critical state, including magnetic 
macroturbulence \cite{tur1,tur2}, kinetic front roughening \cite{front}, 
magnetic avalanches\cite{mav} and dynamic dendritic structures \cite{den1,den2,den3}. 
The latter have been observed both on high-$T_c$ ($YBa_2Cu_3O_7$\cite{den1}) 
and low-$T_c$ (Nb \cite{den2}), superconducting films, and most recently on the newly 
discovered $MgB_2$\cite{den3}. These instabilities are rather 
characteristic of superconductors, besides they also display remarkable similarities with 
other dendritic instabilities in crystal growth \cite{cryst}, nonequilibrium chemical 
and biological systems \cite{rev}, and crack propagation\cite{crack}. This analogy 
is not surprising, because the non isothermal flux dynamics in superconductors is described 
by two nonlinear equations for the magnetic induction 
$B({\bf r},t)$ and temperature $T({\bf r},t)$ somewhat similar to those 
describing dendritic structures in
reaction-diffusion systems \cite{rev}.  

%Motivated by this similarity, 
We performed  numerical 
simulations of coupled equations for the magnetic induction
$B({\bf r},t)$ and temperature $T({\bf r},t)$   and found 
a new mechanism of flux fragmentation in superconductors, which is different from the well-known 
bending instability of moving interface between two phases\cite{cryst}, but 
results from the generic distribution of 
magnetization currents in the critical state.    
Our results give insight into   
unstable flux penetration,
vortex micro avalanches and flux jumps 
in superconductors.    

The effect is best illustrated by 
considering  a slab in a magnetic field $B_0$ (Fig. 1), for which  
distributions of the z-component of magnetic induction, $B({\bf r},t)$, and  
temperature $T({\bf r},t)$ are described by the Maxwell equation coupled to the 
heat diffusion equation.     
	\begin{eqnarray}	%1
	C\partial_tT=\mbox{div}\kappa\nabla T-h(T-T_0)/d+JE(J,T), 
	\label{heat} \\		%2
	\partial_t B=-c\nabla\times{\bf E(J,T)},\quad\quad  {\bf J}=(c/4\pi)\hat{z}\times\nabla B.
	\label{max}
	\end{eqnarray}
Here $C(T)$ is the heat capacity, $\kappa(T)$ is the thermal conductivity, $h(T)$ is the 
heat transfer coefficient to the coolant held at the temperature $T_0$, $d=A/P$, $A$ is the 
area of the sample cross section, $P$ is the perimeter of the cooled sample surface, $B$ is the 
z-component of the magnetic induction, and $E(J,T)$ is the modulus of the electric field, 
which essentially depends on both the local temperature $T({\bf r},t)$ and the current 
density, $J({\bf r},t)=(J_x^2+J_y^2)^{1/2}$.  The spatial 
derivatives are taken with respect to 
x and y, while the term $h(T-T_0)/d$ accounts for
 the surface cooling\cite{add}. 
We will consider the case of high magnetic fields 
$B_0$, much greater then the field of full magnetic flux penetration, $B_p\sim 4\pi J_cd/c$,
for which $E(J,T,B)\approx E(J,T,B_0)$. 

The evolution of $T({\bf r},t)$ and $B({\bf r},t)$ is mostly determined by the 
$E(T,J)$ characteristic, which accounts for  
high-resistive flux flow state at $J>J_c$ and low-resistive flux creep 
state at $J<J_c$, where $J_c$ is the critical current density. The effects considered in this 
work are not very sensitive to the details of $E(J,T)$, so for the numerical simulations of 
Eqs. (\ref{heat})-(\ref{max}) we take the following interpolation formula 
expressed in terms of observable parameters $J_c$, $J_1$, and $\rho$:
	\begin{equation}	%3
	E=\rho J_1\ln [1+\exp (J-J_c)/J_1]
	\label{interp}
	\end{equation}
Here $J_1(T)=\partial J/\partial\ln T$ is the dynamic  
flux creep rate, and $\rho(T)=\rho_n B/B_{c2}$ is the flux flow resistivity.  Below the 
irreversibility field $B<B^*$, where $J_1\ll J_c$,  Eq. (\ref{interp}) reproduces 
the main features of $E(J,T)$ observed in experiment, giving
a linear flux flow dependence $E=(J-J_c)\rho$ for $J>J_c$ and 
the exponential dependence $E=E_c\exp (J-J_c)/J_1$ for $J<J_c$.

The similarity of Eqs. (\ref{heat})-(\ref{max}) with  
generic reaction-diffusion equations\cite{rev}, is due to 
thermal bistability of superconductors\cite{gm}, for which the  
heat balance condition $(T-T_0)h/d=JE(T,J)$ in the right-hand side of Eq. (\ref{heat}) is   
satisfied for 3 different temperatures $T$, as shown in Fig. 1. Here the points 0 and 3  correspond 
to two stable uniforms states: a cold superconducting state with $T\approx T_0$ 
and a hot flux flow/normal state $T_3$ self-sustained by Joule heating. As seen from Fig. 1, 
the bistability occurs, if the current density $J$ exceeds a threshold value $J_m$, for  
which $\rho J_m^2\simeq h(T^*-T_0)/d$. Hence, $J_m\simeq [(T^*-T_0)h/d\rho]^{1/2}$, where 
$T^*(B)$ is the irreversibility temperature at which $J_c(T^*)=0$. 
The superconducting state is unstable with respect to the hot spot 
formation, if $\alpha_s=(J_c/J_m)^2>1$. For typical parameters of HTS films at $T_0=4.2$K: 
$\rho\sim100\mu\Omega$cm, $J_c=10^6 - 10^7 A/cm^2$, $h\sim 1W/cm^2K$, $T^*-T_0\sim 50-100$K 
and $d=1\mu$m, we obtain $\alpha_s\sim 10^2-10^4$, thus the thermal bistability is a  
characteristic feature of both HTS and LTS, especially films because of their higher $J_c$ 
values \cite{gm}.   

We consider the case of weak Joule heating, $T(x,y,t)\simeq T_c$,  for which 
we take into account only the most essential temperature dependence of $E(T)$, while 
$C(T)$, $h(T)$, and $\kappa (T)$ can be taken at $T=T_0$. Then  
Eqs. (\ref{heat})-(\ref{max}) can be written in the following dimensionless form
	\begin{eqnarray}
	\tau{\dot b}=\partial_x[r(j,\theta)\partial_xb]+\partial_y[r(j,\theta)\partial_yb]
	\label{b} \\	
				%1
	{\dot\theta}=\nabla^2\theta-\theta+\alpha j^2r(j,\theta).	
	\label{t}			%2
	\end{eqnarray} 
Here $\theta = (T-T_0)/(T^*-T_0)$, $b=B/B_t$ and $j=[(\partial_xb)^2+(\partial_yb)^2]^{1/2}$ 
are the dimensionless temperature, magnetic field, and current density, respectively, and 
$B_t=4\pi J_1L_h/c$. The  derivatives in 
Eqs. (\ref{b}) and (\ref{t}) are taken with respect to 
normalized time $t/t_h$ and 
coordinates $x/L_h$ and $y/L_h$ measured in the thermal units $t_h=Cd/h$ and $L_h=(d\kappa/h)^{1/2}$.
The evolution of $\theta({\bf r},t)$ and $b({\bf r},t)$ is controlled  by two dimensionless parameters:
	\begin{equation}
	\tau=\frac{4\pi\kappa}{\rho Cc^2},\qquad \alpha=\frac{\rho J_1^2d}{h(T^*-T_0)}
	\label{param}
	\end{equation}
Here $\tau$ is the ratio of the diffusivities of heat and 
magnetic flux, and $\alpha$ quantifies Joule dissipation. For Nb films of Ref. \cite{den2},   
we obtain $\tau\sim 10$ at 4.2K, with $\tau$ rapidly decreasing with increasing $T_0$. 
For HTS at 77K, we get $\tau\sim 10^{-4}-10^{-5}\ll 1$. 
The nonlinear resistivity $r(j,\theta )=\ln [1+\exp (j-j_c(\theta))]/j$ obtained from Eq. (3) 
has asymptotics $r=1-j_c/j$ in the flux flow 
$(j>j_c$) and $r=\exp (j-j_c)/j$  in the flux creep $(j<j_c)$ states, where  
$j_c=J_c(T)/J_1$. We  linearize $j_c=j_0(1-\theta)$ around $T_0$, neglecting the temperature 
dependencies of $J_1$, $\rho$ \cite{temp}. 

We performed 2D numerical simulations of Eqs. (\ref{b}) and (\ref{t}) to calculate propagation 
of a  magnetic hotspot  of resistive phase across a superconductor (see Fig. 1). 
The process is initiated by a local heat pulse applied to the sample surface, which  
models the experiment by Leiderer et al., \cite{den1}, who triggered the magnetic 
dendrite instability by a laser pulse. To address the effect of material inhomogeneities, 
we considered both uniform superconductors with 
$J_c$ independent of spatial coordinates and nonuniform superconductors with the 
critical current density periodically modulated over macroscopic scales $2\pi/k$, much 
larger than the spacing between flux lines,  
$J_c(x,y,T)=J_{c0}(T)[1+\epsilon\sin(kx)\sin(ky)]$ with $\epsilon < 1$. 
The latter case also models superconducting films with periodic arrays of holes, which have 
recently attracted much interest \cite{hole,vlasko1}.  The resulting evolutions of the temperature 
distributions shown in Figs. 2 and 3  display a rather striking behavior 
described below. 

For $\tau\ll 1$, the heat pulse applied to a uniform superconductor 
triggers a hot spot propagation across the sample, as shown in Fig. 2a. 
Such propagation appears stable until the hot domain reaches the center of the sample, 
where magnetization currents change direction.  Then the resistive domain undergoes a 
cascade of successive splittings into alternating stripes of low and high electric 
fields and temperatures. After each successive splitting, the part of the resistive domain near 
the central line cools down, but then the hot filaments of resistive state 
start propagating again from the upper part of the resistive domain through the 
preceding dendritic structure toward the central line. 
However, each time the hot filaments cross the central line, 
they split again, causing 
new dendrites of alternating low and high-J filaments to grow as
shown in  Figures 2b and 2c. 
Eventually the Joule dissipation causes 
the electric field in the hot dendritic structures to decay below the threshold, at which 
further hot spot propagation becomes impossible, thus a frozen entangled pattern of  
current filaments forms.    

As follows from Figures 2d-2f, spatial inhomogeneities in $J_c$ can bring  
about new features of the hot spot propagation, causing fragmentation of 
the hot spot  even before it crosses the central line. The periodic modulation of 
$J_c(x,y)$ also gives rise to additional side branching and preferential flux propagation 
along the nearest neighbor directions at angles $\pm 45^{\circ}$, causing 
further interconnection of neighboring hot spot branches. This effect is similar to that 
observed by Vlasko-Vlasov {\it et al} on superconducting films with periodic arrays of holes and 
magnetic dots\cite{vlasko1}. For $\tau\ll 1$, the diffusion of magnetic flux in the 
normal state occurs much faster than the heat diffusion. Therefore, magnetic flux rapidly 
penetrates the hot regions of the filamentary current structure in Fig. 2, forming 
dendritic flux front patterns reminiscent of those observed in magneto-optical 
experiments.\cite{den1,den2,den3} 

The flux fragmentation can be described as follows.  
For $\tau\ll 1$,  the electric field $\bf{E}$ becomes nearly potential, 
$\nabla\times\bf{E}\approx 0$, thus the magnetization currents tend to bypass propagating hot 
resistive domain, similar to that of a steady-state current flow past a planar obstacle \cite{ag}. 
Thus,  the current density in the resistive domain decreases, forcing the excess magnetization 
currents to flow along the domain interface. The high interface currents cause 
strong local enhancement of electric field and dissipation, widening the resistive domain 
near its end accelerating the propagation velocity. At the same time, the temperature 
in the center of the resistive domain decreases, facilitating recovery of the superconducting 
state when the domain crosses the central line, 
where magnetization currents change direction. In this region the interface 
currents at the bottom part of the domain are partly compensated by the opposite 
magnetization currents, which strongly reduces Joule dissipation and stops hot 
spot propagation. As a result, a triangular region at the tip of the resistive domain becomes 
superconducting, and then the process repeats as described above.

For $\tau\gg 1$, the hot spot propagation occurs at the frozen 
distribution of magnetic fields and currents, and   
the hot spot dynamics changes as shown in Fig. 3. Instead of the multiple splitting 
characteristic of $\tau\ll 1$, the heat pulse at $\tau\gg 1$ first initiates stable propagation 
of a resistive hot spot, then 
its splitting into two parts which move apart and eventually disappear.  This behavior 
occurs if the energy of the heat pulse Q is below the critical value $Q_c$. For 
$Q>Q_c$, the heat pulse creates a larger hotspot which then expands and  
propagates over the entire sample\cite{gm}.

The dendritic flux penetration described above can also 
be regarded as a {\it micro avalanche} of a macroscopic bundle of vortices, which does not 
trigger a global flux jump instability or thermal quench of the whole sample. Such  micro avalanches 
only cause local transient temperature spikes, leaving behind the frozen 
flux dendrite structures shown in Figure 4. Each micro avalanche thus 
results in a partial flux penetration, which reduces the total 
magnetic moment of a sample and manifests itself in steps on magnetization 
curves $M(B)$. Notice that micro avalanches in increasing magnetic field may also be 
initiated by surface defects (regions with lower $J_c$), which are sources of excess 
steady-state Joule dissipation. Such common defects, which have been revealed 
by magneto-optical imaging of HTS\cite{mo}, can trigger both global flux jumps\cite{fj} and 
local vortex micro avalanches.  The micro avalanches cause steps in $M(B)$ as observed 
on Nb films at low temperatures \cite{mav}. After many microavalanches the critical state 
eventually turns into a frozen "turbulent" current structure built of individual 
dendritic fragments, like that in Fig. 4      

The results this work may capture the essential physics of  
dendritic flux instability observed on $YBa_2Cu_3O_7$ \cite{den1}, Nb \cite{den2} and 
$MgB_2$\cite{den3} films,  
although, for a more quantitative comparison of the theory with experiment, other factors 
should also be taken into account. Indeed,  the experiments \cite{den1,den2,den3} correspond to 
thin films in low perpendicular magnetic fields, which requires the account of the 
nonlocal flux diffusion\cite{ehb} and the geometrical barrier\cite{zeld}, 
whereas our model describes a slab in high parallel magnetic 
fields unaccessible to magneto-optical technique. Another intriguing result discovered in 
Ref. \cite{den1} is a superfast dendritic flux propagation with the velocities exceeding 
the speed of sound $c_s$. This also requires 
invoking additional mechanisms, because the thermal 
velocities of hotspot propagation $v\sim J[\kappa\rho /(T^*-T_0)]^{1/2}/C$ \cite{gm} 
is smaller than $c_s$.
The superfast dendritic flux propagation 
might be due to electron overheating, so that an equation similar to Eq. (1) 
actually describes the electron temperature, higher than the lattice temperature $T_0$. 
Then the lattice heat capacity C in the above estimate for $v(J)$ is to be replaced by a much smaller 
electron heat capacity, which increases $v(J)$ by 1-2 orders of 
magnitude. This situation may occur at low T, if the time of the electron-phonon energy 
relaxation becomes larger than the thermal time $t_h$.\cite{rt}. 

In conclusion, we proposed a new mechanism of magnetic flux fragmentation in 
superconductors. The instability manifests itself as 
a vortex micro avalanche, accompanied by a transient local Joule dissipation and 
eventually resulting in a frozen multi-filamentary structure of magnetization currents. 
These effects give rise to dendritic flux penetration into superconductors, partial flux jumps 
and steps on magnetization curves.      

This work was supported by the NSF  MRSEC (DMR 9214707) (AG),
U.S. Department of Energy, BES-Materials Sciences
(\# W-31-109-ENG-38) (IA and VV). We are grateful to Peter Kes and Wim van Saarloos for hospitality  
during the initial stage of this work and to Vitalii Vlasko-Vlasov for illuminating discussions.

\vspace{0.15cm}
\begin{figure}[h]
\centerline{ \psfig{figure=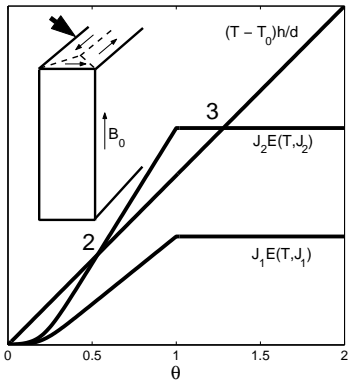,height=2in}}
\vspace{0.8cm}

\caption{Graphic solution of the heat 
balance condition $(T-T_0)h/d=JE(\theta,J)$ in a  
superconductor. The power of the Joule heat release  
$JE(J,\theta)$ is plotted as a function of T for  
$J_1<J_m$  and $J_2>J_m$, where $E(J,\theta)$  is given by 
Eq. (\protect \ref{interp}), 
and $\theta = (T-T_0)/(T^*-T_0)$. 
Inset shows the sample geometry, magnetic field is parallel to $z$ 
axis, $x$-axis is directed along the sample surface and $y$ is 
perpendicular to the sample surface.  
The local heat pulse (indicated as big arrow) 
triggers the magnetic hotspot propagation 
across magnetization currents presented in detail in Figs. 2 and 3.
}
\label{Fig1}
\end{figure}

\begin{figure}[h]
\centerline{ \psfig{figure=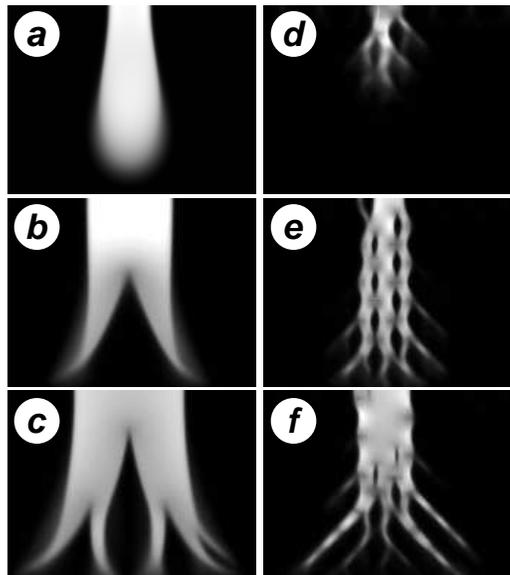,height=3in}}
\caption{
Gray-coded dynamic temperature maps $\theta({\bf r},t)$ 
for the flux fragmentation instability in 
homogeneous system for $t/t_h=$8 (a), 20 (b), 34 (c)
at $\alpha=0.008$, $\tau^{-1}=150$ and 
the initial current in the Bean state, 
$j_0=18$ (white corresponds to $\theta>1$, and black to $\theta=0$). 
Domain of integration: $-150 L_h<y<150 L_h$, 
$0<x<600 L_h$, periodic
boundary conditions in $x$ direction; no-flux $\partial \theta/\partial y$ 
at  $y=\pm 150 L_h$ for the temperature and $b=const$ 
for the magnetic field. 
Each panel shows the upper half of the sample and $1/3$ of total lenght:
$0<y<150  L_h$, $200 L_h < x <400 L_h$. 
The magnetization currents change direction on the bottom part of each panel,  
i.e for $y=0$. 
Same for the system with periodic modulation in $J_c(x,y)$ for $2\pi/k=30 L_h$, 
$\epsilon = 0.5 $, 
$t=$6(d), 16(e),  28(f) and $j_0=20$. }
\label{Fig2}
\end{figure}

\begin{figure}[h]
\centerline{ \psfig{figure=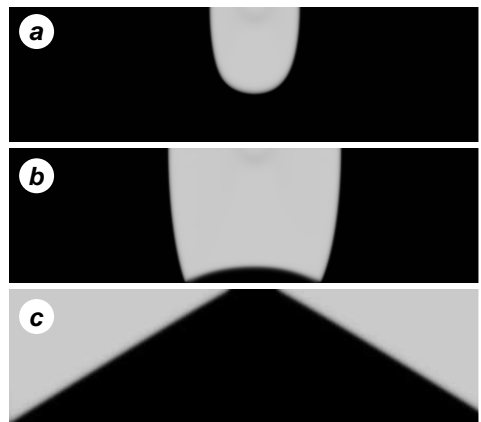,height=2.25in}}
\caption{
Same dynamic temperature maps as in Figure 2 for $\tau=1$, $\alpha=0.006$, 
$j_0=18$, and  $t/t_h=$8 (a), 20 (b), 34 (c).
}
\end{figure}

\begin{figure}[h]
\centerline{ \psfig{figure=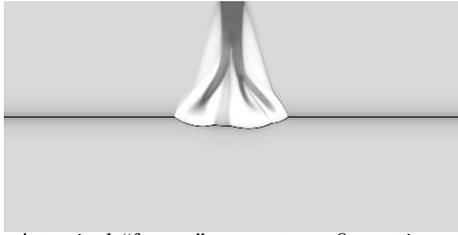,height=1.2in}}
\caption{A typical ``frozen'' current configuration after completion of the 
fragmentation for $\tau^{-1}=600$. The gray shades change from black ($J=0$) to
white (maximum $J$). Full domain of integration is shown.     
Black line in the middle is due to reverse of direction of 
the magnetization current.  
}
\label{Fig4}
\end{figure}


\begin{references}
\bibitem{add} The term $h(T-T_0)/d$ accounts for the heat removal from 
%a sample for
%: 1. a film of thickness 
%$d\ll \kappa/h$ with the boundary condition 
%$\kappa\partial_zT=-(T-T_0)h$ at the surface of a substrate, 
%so that Eq. (\protect\ref{heat}) describes 
%$T(x,y,t)$ averaged over the film thickness;  
a slab in a parallel field, for which this term provides 
the correct integral heat balance, $(T-T_0)h/d = JE(T)$, if  
we adopt the effective boundary conditions 
$\partial_y T=0$ at the surface.  Our qualitative results are 
not sensitive to the thermal boundary conditions, because 
the dendritic flux instability occurs in a central region of the sample,  and  
the velocity of the hotspot propagation for 
$\alpha_s\gg 1$ is independent of $h$\cite{gm}.
\end{references}
\end{document}